\documentclass[a4paper, prl, twocolumn,superscriptaddress,showpacs]{revtex4}

\pdfoutput=1
\usepackage{amssymb}
\usepackage{amsmath}
\usepackage{epsfig}
\usepackage{color}
\usepackage{graphics, graphicx}
\usepackage{bbold}
\usepackage{psfrag}
\usepackage{mathcomp}
\usepackage{subfigure}
\usepackage{verbatim}
\usepackage{color}
\usepackage[colorlinks,citecolor=blue]{hyperref}

\begin{document}

\date{\today}
\pacs{03.75.Ss, 03.75.Lm, 05.30.Fk}
\title{Molecule and polaron in a highly polarized two-dimensional Fermi gas with spin-orbit coupling}

\author{Wei Yi}
\email{wyiz@ustc.edu.cn}
\affiliation{Key Laboratory of Quantum Information, University of Science and Technology of China,
CAS, Hefei, Anhui, 230026, People's Republic of China}
\author{Wei Zhang}
\email{wzhangl@ruc.edu.cn}
\affiliation{Department of Physics, Renmin University of China, Beijing 100872, People's Republic of China}

\begin{abstract}
We show that spin-orbit coupling (SOC) gives rise to pairing instability in a highly polarized two-dimensional Fermi gas for arbitrary interaction strength. The pairing instability can lead to a Fulde-Ferrell-Larkin-Ovchinnikov (FFLO)-like molecular state, which undergoes a first-order transition into a pairing state with zero center-of-mass momentum as the parameters are tuned. These pairing states are metastable against a polaron state dressed by particle-hole fluctuations for small SOC. At large SOC, a polaron-molecule transition exists, which suggests a phase transition between the topological superfluid state and the normal state for a highly polarized Fermi gas in the thermodynamic limit. As polarization in a Fermi gas with SOC is induced by the effective Zeeman field, we also discuss the influences of the effective Zeeman field on the ground state of the system. Our findings may be tested directly in future experiments.
\end{abstract}
\maketitle


Spin-orbit coupling (SOC), a non-Abelian gauge field, has been shown to play a fundamentally important role in many interesting systems in condensed matter physics, e.g., topological insulators \cite{kanereview}, quantum spin Hall materials \cite{xiaoreview}, etc. The recent development of a synthetic gauge field, SOC in particular, in ultracold atoms has stimulated a tremendous amount of interest in the study of the effects of SOC within these systems \cite{gauge1,gauge2exp,zhairev}. By breaking the inversion symmetry, the SOC may induce novel quantum phases, e.g., the unconventional superfluidity in an ultracold Bose gas \cite{wucongjun,zhaibec}, or the topological superfluid phase (TSF) in a polarized Fermi gas with $s$-wave pairing order \cite{zhang,sato,soc1,melo,helianyi,huhui}. Notably, in the latter case, Majorana zero modes can be stabilized at the center of vortex excitations, which may be used as resources for topological quantum computation \cite{majmode,tqc2}.

For a polarized two-dimensional (2D) Fermi gas with SOC, a peculiar behavior is the existence of the pairing instability in the large polarization limit \cite{soc1,wy2d,duan}. This is in stark contrast to the case without SOC, where the pairing state becomes unstable against a normal gas beyond the so-called Chandrasekhar-Clogston limit \cite{clogston}. The persistence of the pairing instability can be attributed to the breaking of inversion symmetry, which modifies the topology of the Fermi surface and renders singlet $s$-wave pairing possible in the large polarization limit \cite{soc1}. This can lead to the interesting scenario where pairing exists in the presence of a single Fermi surface, i.e., when the chemical potential lies in the gap between the two helicity bands. In this case the system is in the TSF phase. For a uniform 2D gas, the TSF phase is stable in the large polarization limit on the mean-field phase diagram \cite{xiaosen}. However, it has been shown theoretically \cite{pethick,parish} and demonstrated very recently in experiments \cite{koehlexp}, that in the absence of SOC the ground state of a highly polarized 2D Fermi gas is a polaron state in the weak-coupling limit, i.e., an impurity atom dressed by particle-hole fluctuations of the Fermi sea. In the presence of SOC, naturally one expects the interplay of SOC, pairing, and fluctuation leads to rich physics. In particular, it is interesting to study the stability of the TSF phase against a normal state with particle-hole fluctuations in a highly polarized Fermi gas.

In this work, we investigate a spin-orbit coupled 2D Fermi gas in the large polarization limit. In the presence of SOC, the spin polarization can be induced by an effective Zeeman field, which is tuned by adjusting the laser parameters in a typical scheme for synthetic SOC \cite{gauge1,gauge2exp}. To model the ground state of the system, we adopt variational ansatz states following Refs. \cite{pethick,parish,chevy,combescot}, which effectively project the wave functions into the subspace of the large polarization limit. We then study in detail the properties of both the molecular state and the polaron state in the presence of an effective Zeeman field. We find that under appropriate effective Zeeman fields the SOC-induced pairing instability leads to a Fulde-Ferrell-Larkin-Ovchinnikov (FFLO)-like pairing state \cite{fflo} with nonzero center-of-mass momentum in the weak coupling limit. The FFLO-like pairing state can undergo a first-order transition into a pairing state with zero center-of-mass momentum as SOC increases or as the interaction is tuned. For small SOC, we find that the energy of a polaron state is always lower than that of the molecular pairing state. However, a polaron-molecule transition exists for sufficiently large SOC. This suggests a phase transition between the normal state and the TSF state in the thermodynamic limit. Finally, we show that the boundaries between the different states can be shifted as the effective Zeeman field increases. With progress in the experimental investigation of 2D Fermi gases \cite{koehlexp,2dgasexp} and the recent realization of SOC in a degenerate Fermi gas \cite{zhangexp}, our study has interesting implications for future experiments.

\emph{Model}.--
We consider the system in the large polarization limit induced by an effective Zeeman field $h$. In the presence of Rashba SOC, the Hamiltonian can be written as
\begin{align}
H&=\sum_{\mathbf{k},\sigma}\epsilon_{\mathbf{k}}c^{\dag}_{\mathbf{k},\sigma}c_{\mathbf{k},\sigma}+\frac{U}{\cal S}\sum_{\mathbf{k},\mathbf{k}',\mathbf{q}} c^{\dag}_{\mathbf{k},\uparrow}c^{\dag}_{\mathbf{k}',\downarrow}c_{\mathbf{k}'+\mathbf{q},\downarrow}c_{\mathbf{k}-\mathbf{q},\uparrow}\nonumber\\ &+\sum_{\mathbf{k}}\left(\alpha e^{i\varphi_{\mathbf{k}}} k c^{\dag}_{\mathbf{k},\uparrow}c_{\mathbf{k},\downarrow}+\alpha e^{-i\varphi_{\mathbf{k}}}kc^{\dag}_{\mathbf{k},\downarrow}c_{\mathbf{k},\uparrow}\right)\nonumber\\
&+h\sum_{\mathbf{k}}a_{\mathbf{k},\downarrow}^{\dag}a_{\mathbf{k},\downarrow},
\end{align}
where $\epsilon_{\mathbf{k}}=\hbar^2k^2/2m$, the pseudospin of the atoms $\sigma=(\uparrow,\downarrow)$, $\alpha$ is the SOC strength, $\varphi_{\mathbf{k}}=\arg{\left(k_x+ik_y\right)}$, and ${\cal S}$ is the quantization volume in two dimensions. Here we only consider the case where atoms of different spin species have the same mass $m$. The bare $s$-wave interaction rate $U$ should be renormalized following the standard relation in two dimensions \cite{renorm},
\begin{equation}
\frac{1}{U}=-\frac{1}{\cal S}\sum_{\mathbf{k}}\frac{1}{E_b+2\epsilon_{\mathbf{k}}},
\end{equation}
where $E_b$ is the binding energy of the two-body bound state in two dimensions in the absence of SOC. For cold atom systems, this two-body binding energy can be tuned, for instance, via the Feshbach resonance technique.

\emph{Molecular state}.--
We first investigate the pairing instability in the presence of SOC, and the properties of the resulting molecular state. Consider a variational ansatz of molecular state with center-of-mass momentum $\mathbf{Q}$:
\begin{align}
|M\rangle&=\sum_{\mathbf{k}}\left(\phi^{\uparrow\downarrow}_{\mathbf{k}}c^{\dag}_{\mathbf{Q}-\mathbf{k},\downarrow}c^{\dag}_{\mathbf{k},\uparrow} +\phi^{\uparrow\uparrow}_{\mathbf{k}}c^{\dag}_{\mathbf{Q}-\mathbf{k},\uparrow}c^{\dag}_{\mathbf{k},\uparrow}\right.\nonumber\\
&\left.+\phi^{\downarrow\downarrow}_{\mathbf{k}} c^{\dag}_{\mathbf{Q}-\mathbf{k},\downarrow}c^{\dag}_{\mathbf{k},\downarrow}\right)|N-1\rangle,\label{molstate}
\end{align}
where $|N-1\rangle$ represents a Fermi sea with $N-1$ spin-up atoms. Due to the SOC, we now have triplet-pairing components $\left\{\phi^{\uparrow\uparrow}_{\mathbf{k}},\phi^{\downarrow\downarrow}_{\mathbf{k}}\right\}$ in addition to the singlet-pairing wave function $\phi^{\uparrow\downarrow}_{\mathbf{k}}$. The momentum subscripts in the wave functions are constrained by Pauli blocking, such that $k>k_F$ for $\phi^{\uparrow\downarrow}_{\mathbf{k}}$ and $\phi^{\uparrow\uparrow}_{\mathbf{k}}$, and $|\mathbf{Q}-\mathbf{k}|>k_F$ for $\phi^{\uparrow\uparrow}_{\mathbf{k}}$, where $k_F$ is the Fermi wave vector given by $E_F=\hbar^2k^2_F/2m$, and $E_F$ is the Fermi energy of spin-up fermions. To focus on the properties of the pairing instability, we neglect the particle-hole fluctuations here and only consider the `bare' molecular state in Eq. (\ref{molstate}). Note that by taking the ansatz in Eq. (\ref{molstate}), we have effectively projected the ground state into a subspace that corresponds to the large polarization limit where few spin-down atoms coexist with a polarized Fermi sea of spin-up atoms. Terms with further spin flips are suppressed by the effective Zeeman field $h$ and are projected out. Here, to be of more experimental relevance, we fix the effective Zeeman field, which typically corresponds to fixing the laser parameters for synthetic SOC. As a result, the population of the spin-down atoms in the ground state fluctuates slightly around unity, given appropriately chosen Zeeman field strengths.

Minimizing the functional $\left\langle M|\left(H-E_M\right)|M\right\rangle$, we get a self-consistent equation for the ground state energy $E_M$ of the molecular state, relative to the Fermi sea of $N$ spin-up atoms \cite{seesupp}:
\begin{align}
\frac{1}{U}&=\frac{1}{\cal S}\sum_{k>k_F} \bigg[ E_M+E_F-h-\left(\epsilon_{\mathbf{k}}+\epsilon_{\mathbf{Q}-\mathbf{k}}\right)\nonumber\\ &-\frac{2\alpha^2k^2}{E_M+E_F-2h-\left(\epsilon_{\mathbf{k}}+\epsilon_{\mathbf{Q}-\mathbf{k}}\right)}\nonumber\\ &-\theta(|\mathbf{Q}-\mathbf{k}|-k_F)\frac{2\alpha^2\left|\mathbf{Q}-\mathbf{k}\right|^2} {E_M+E_F-\left(\epsilon_{\mathbf{k}}+\epsilon_{\mathbf{Q}-\mathbf{k}}\right)}\bigg]^{-1},\label{molenergyeqn}
\end{align}
where $\theta(x)$ is the Heaviside step function. To find the molecular ground state, we should further minimize the energy solved from Eq. (\ref{molenergyeqn}) with respect to the center-of-mass momentum $\mathbf{Q}$. In the absence of SOC, the center-of-mass momentum of the molecular state is zero for large $E_b$, becomes finite, i.e. FFLO-like, at $E_b=2E_F$ through a second-order transition, and approaches the Fermi wave vector $k_F$ at $E_b=0.5E_F$ where the bound state merges into the continuum \cite{pethick}. Importantly, there is no pairing instability for $E_b<0.5E_F$ without SOC in the large polarization limit. We will see that this simple picture is drastically modified by SOC.

\begin{figure}[tbp]
\includegraphics[width=8.5cm]{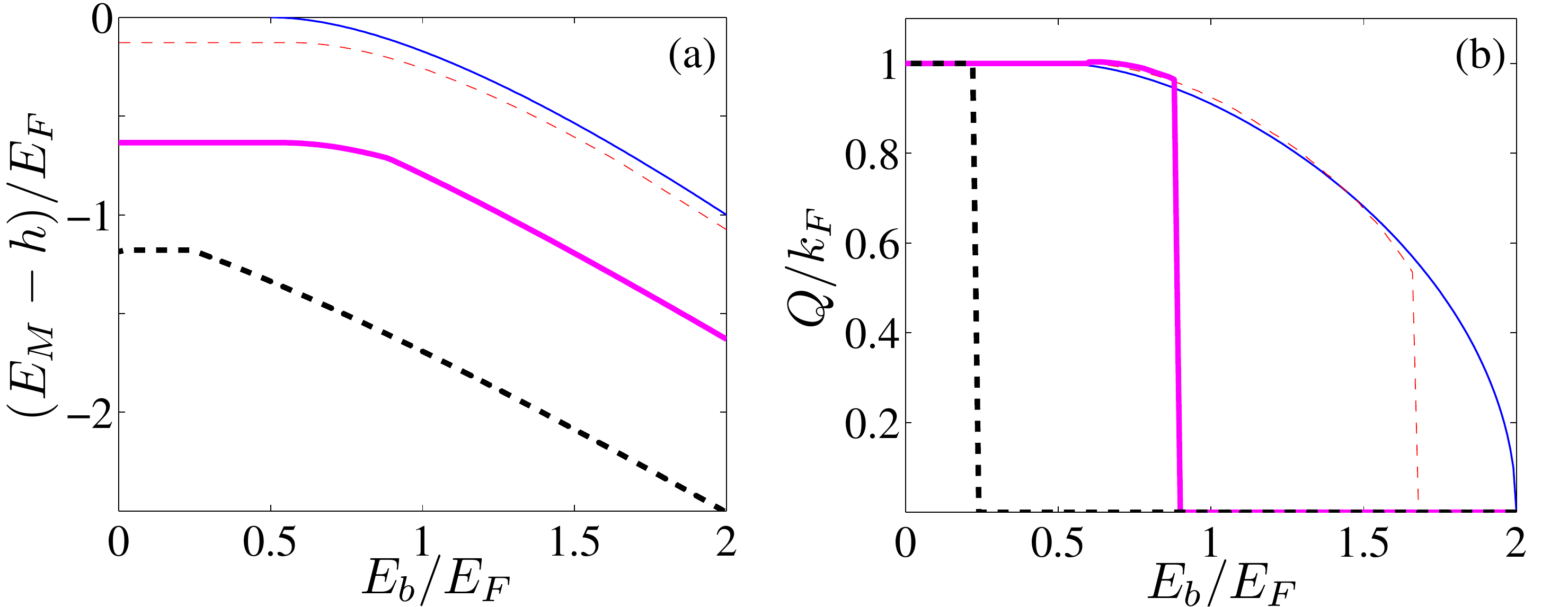}
\caption{ (Color online) (a) The molecular energy $(E_M-h)/E_F$ as a function of $E_b/E_F$ for different SOC strengths and a fixed effective Zeeman field $h/E_F=0.5$: $\alpha k_F/E_F=0$ (thin solid  line), $\alpha k_F/E_F=0.2$ (thin dashed line), $\alpha k_F E_F=0.6$ (bold solid line), $\alpha k_F/E_F=1$ (bold dashed line). (b) Evolution of the center-of-mass momentum of the molecular ground state for different SOC strengths and a fixed effective Zeeman field $h/E_F=0.5$. }\label{figEbvsE}
\end{figure}

We numerically solve Eq. (\ref{molenergyeqn}) and minimize the solution with respect to the center-of-mass momentum $Q$ for a fixed Zeeman field $h=0.5E_F$. The evolutions of the molecular energy as well as the center-of-mass momentum $Q$ are shown in Fig. \ref{figEbvsE}. Compared to the case without SOC, an outstanding difference is that the pairing instability persists into the weak-coupling limit even with infinitesimally small SOC. The bound state energy approaches an SOC-dependent asymptotic value in the weak-coupling limit and never crosses the continuum threshold $E_M=h-\alpha^2k_F^2/2E_F$. This is consistent with the previous many-body calculations, where the superfluid order parameter does not vanish for any polarization in the presence of SOC \cite{soc1,wy2d,duan}. The pairing instability in the large polarization limit is actually a consequence of the combined effects of the SOC and the effective Zeeman field. While SOC breaks the inversion symmetry and mixes the spins into different helicity bases, the effective Zeeman field breaks the time reversal symmetry and opens a gap in the energy spectrum. As the spins are mixed in the lower branch, $s$-wave pairing is possible even in the large polarization limit \cite{soc1}.  Numerically, the pairing instability shows up as a singularity in Eq. (\ref{molenergyeqn}) when $E_b\rightarrow 0$, similar to the behavior of the gap equation in the many-body case. From the many-body mean-field calculations, the ground state of a uniform 2D Fermi gas is a TSF phase in the large polarization limit \cite{xiaosen}. Apparently, the pairing state that we discuss here is related to the topological superfluid phase in the thermodynamic limit when a finite density of spin-down atoms are present \cite{wy2d}.

\begin{figure}
\includegraphics[width=8.5cm]{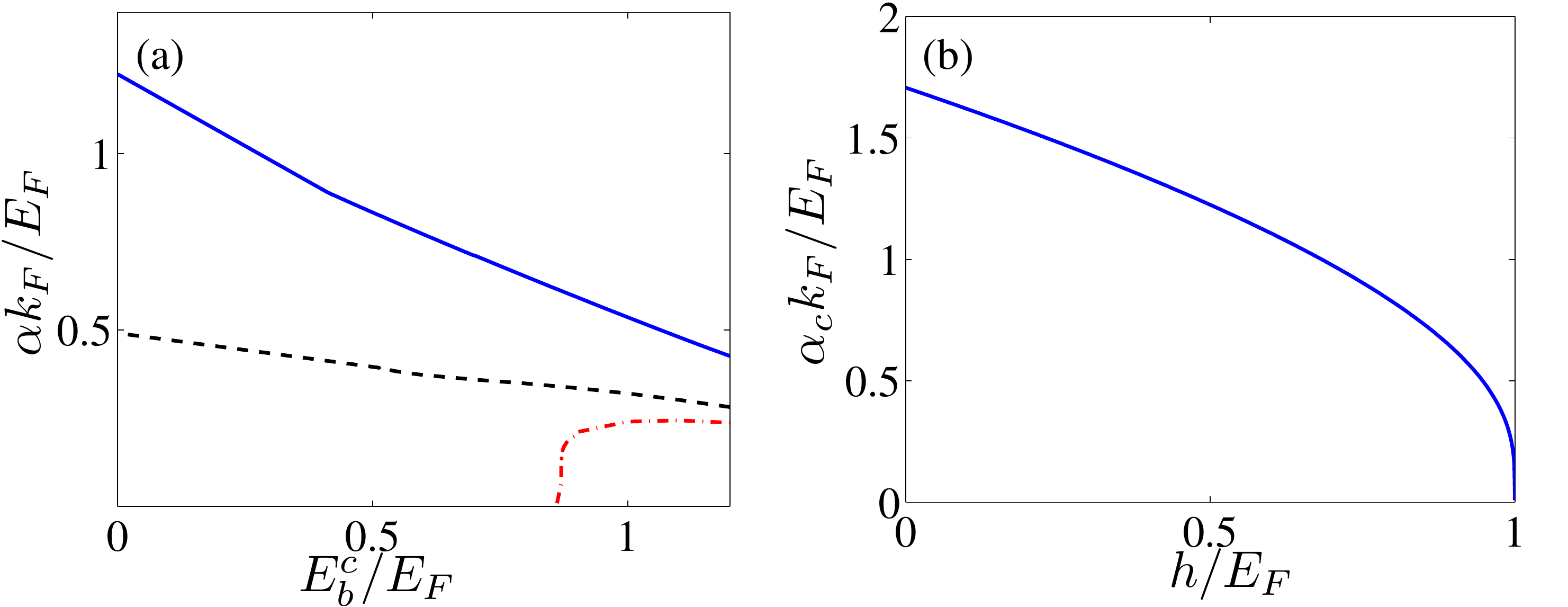}
\caption{(Color online) (a) The critical $E_b^c$ for the first-order transition to FFLO-like pairing state as a function of the SOC strength for various effective Zeeman fields: $h/E_F=0.5$ (solid line), $h/E_F=0.95$ (dashed line), $h/E_F=1.1$ (dash-dotted lines). (b) The critical SOC strength in the weak-coupling limit $E_b=0$ as a function of the effective Zeeman field $h$.}
\label{figcritalp}
\end{figure}

Another important observation is that, for small SOC, the pairing instability leads to an FFLO-like pairing state with a finite center-of-mass momentum  ${\bf Q}$ whose magnitude approaches the Fermi wave vector $k_F$ in the weak-coupling limit. This is shown in Fig. \ref{figEbvsE}(b). As the interaction is tuned toward the strongly interacting region, the center-of-mass momentum becomes smaller and drops to zero at a critical $E_b^c$, where the system undergoes a first-order transition to a pairing state with zero center-of-mass momentum. The critical $E_b^c$ for this first-order transition is, in general, a function of the SOC strength and the effective Zeeman field $h$. As demonstrated in Fig. \ref{figcritalp}(a), for small $h$, $E_b^c$ decreases monotonically as the SOC strength becomes larger, and eventually vanishes at a critical SOC strength $\alpha_c$. Beyond $\alpha_c$ only pairing states with zero center-of-mass momentum are stable for arbitrary interaction strength. For larger $h$, the dependence of $E_b^c$ on the SOC strength becomes non-monotonic. This suggests that the pairing state becomes non-FFLO-like in the weak coupling limit for large $h$, and that one may encounter up to two first-order transitions if the interaction is tuned with fixed SOC and Zeeman field. Hence for a fixed Zeeman field, the stability of the FFLO state is affected by SOC in two contrasting ways: while the competition between the FFLO state and the pairing state with zero center-of-mass momentum limits the stability region of the FFLO state, the pairing instability induced by SOC effectively increases the stability region of the FFLO state, especially in the weak-coupling regime. An outstanding question here is what are the topological properties of the FFLO pairing state, which deserves further investigation in the future.

Now that we understand the nature of the pairing states in the weak-coupling limit, we may derive an analytical expression for the molecular energy in this limit. Similar to the divergence of the gap equation in the many-body case \cite{soc1,wy2d}, the summation in Eq. (\ref{molenergyeqn}) must diverge when $E_b\rightarrow 0$, as dictated by the pairing instability. The energy of the molecular ground state in the weak-coupling limit corresponds to the lowest-lying singularity of the summand on the right-hand side of Eq. (\ref{molenergyeqn}). To evaluate this energy, we set the denominator in the summand to zero and minimize the molecular energy with the constraint $k>k_F$. As the center-of-mass momentum $Q$ is either $k_F$ or $0$ in the weak-coupling limit, we need only consider these two cases. For the FFLO state with $Q=k_F$, we find the lowest energy for the occurrence of the singularity at $E^f_M=(3h-\sqrt{h^2+8\alpha^2 k_F^2})/2$, with $\left\{\varphi\rightarrow 0, k\rightarrow k_F\right\}$ ($\varphi$ is the angle between ${\bf Q}$ and ${\bf k}$). For $Q=0$, the asymptotic energy is given by $E^f_M=h-E_F-(h^2+\alpha^4k_F^4/E_F^2)E_F/2\alpha^2k_F^2$ for $\alpha^4k_F^4/E_F^2-h^2>4\alpha^2k_F^2 $, and $E^f_M=E_F+h-\sqrt{h^2+4\alpha^2k_F^2}$ for $\alpha^4k_F^4/E_F^2-h^2<4\alpha^2k_F^2 $. With these, the critical SOC strength $\alpha_c$ at which the FFLO-like pairing state vanishes at $E_b=0$ can be calculated as a function of the effective Zeeman field $h$ [see Fig. \ref{figcritalp}(b)]. Apparently, the stability region of the FFLO-like pairing state decreases as the Zeeman field $h$ increases. In particular, the FFLO-like pairing state vanishes in the weak coupling limit beyond $h/E_F=1$ for arbitrary SOC .

\begin{figure}
\includegraphics[width=8cm]{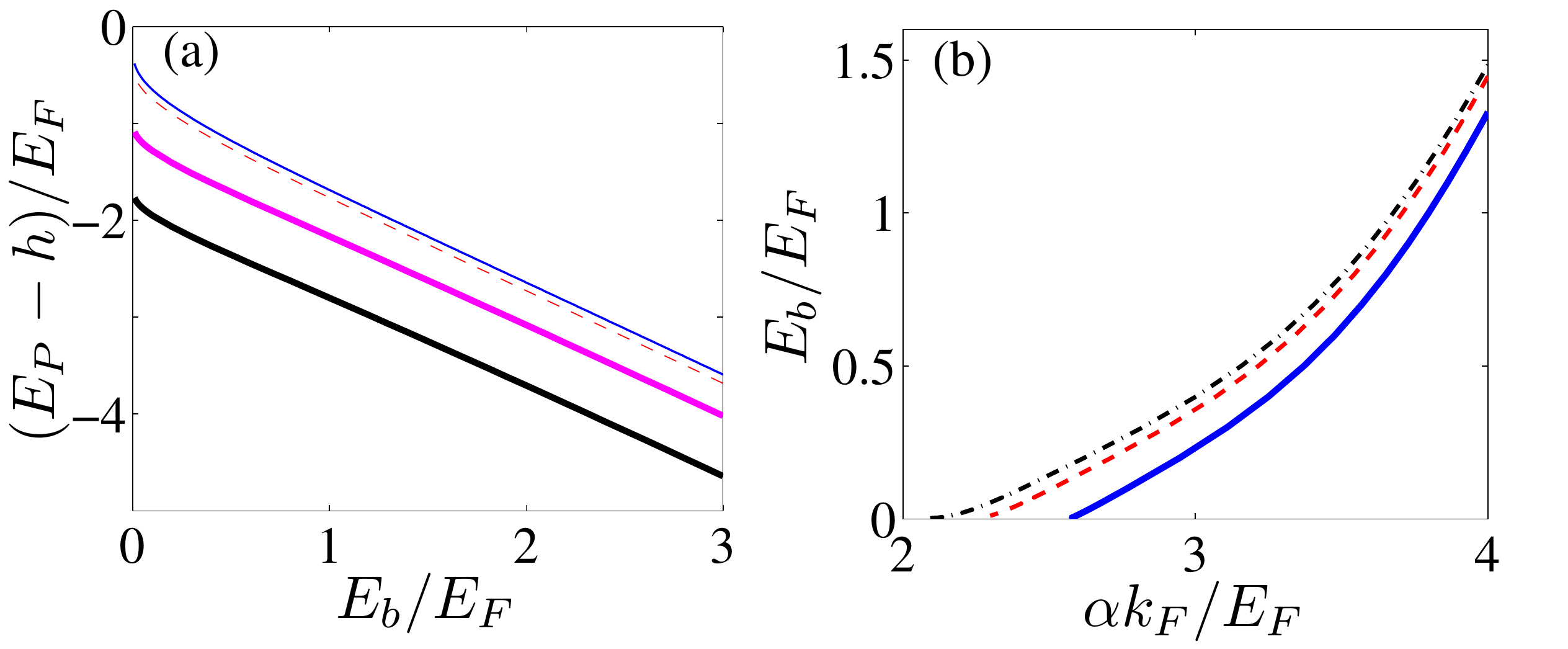}
\caption{(Color online) (a) The energy of the polaron state as a function of $E_b/E_F$ for various SOC strengths with fixed effective Zeeman field $h/E_F=0.5$: $\alpha k_F/E_F=0$ (thin solid line), $\alpha k_F/E_F=0.2$ (thin dashed line), $\alpha k_F E_F=0.6$ (bold solid line), $\alpha k_F/E_F=1$ (bold dashed line). (b) The polaron-molecule transition boundary on the $\left\{E_b/E_F,\alpha k_F/E_F\right\}$ plane for various Zeeman fields: $h/E_F=0.5$ (solid line), $h/E_F=0.95$ (dashed line), $h/E_F=1.1$ (dash-dotted line).}\label{figpolaron}
\end{figure}

\emph{Polaron state}.--
The molecular states that we have discussed are the ground state of the system if only the `bare' polaron state $|P\rangle=\psi_0c^{\dag}_{0\downarrow}|N\rangle$ is considered \cite{parish}. In two dimensions, we should include the particle-hole fluctuations above this `bare' polaron state for a more realistic calculation. The energy of the dressed polaron state is then lowered and the molecular state in Eq. (\ref{molstate}) may no longer be the ground state. Indeed, in the absence of SOC, it has been shown previously that for a single impurity atom in the presence of a polarized Fermi sea in two dimensions, a polaron state dressed by particle-hole fluctuations has lower energy than the `bare' molecular state that is not dressed by such fluctuations \cite{pethick,parish}.

We consider a variational ansatz for such a polaron state with one particle-hole density fluctuation in the large polarization limit
\begin{align}
|P\rangle&=\psi_0c^{\dag}_{0\downarrow}|N\rangle+\sum_{k>k_F,q<k_F}\psi^{\uparrow\downarrow}_{\mathbf{k},\mathbf{q}}c^{\dag}_{\mathbf{q}-\mathbf{k},\downarrow}c^{\dag}_{\mathbf{k},\uparrow} c_{\mathbf{q},\uparrow}|N\rangle\nonumber\\
&+\sum_{\mathbf{k},q<k_F}\psi^{\downarrow\downarrow}_{\mathbf{k},\mathbf{q}}c^{\dag}_{\mathbf{q}-\mathbf{k},\downarrow}c^{\dag}_{\mathbf{k},\downarrow} c_{\mathbf{q},\uparrow}|N\rangle\nonumber\\
&+\sum_{q<k_F,k>k_F}\theta(|\mathbf{q}-\mathbf{k}|-k_F)\psi^{\uparrow\uparrow}_{\mathbf{k},\mathbf{q}}c^{\dag}_{\mathbf{q}-\mathbf{k},\uparrow}c^{\dag}_{\mathbf{k},\uparrow}c_{\mathbf{q},\uparrow}|N\rangle.
\end{align}
Similar to the molecular case, the effective Zeeman field is applied to induce the spin polarization. Again, we fix the Zeeman field in the following for simplicity and for experimental relevance.

Minimizing $\left\langle P|\left(H-E_P\right)|P\right\rangle$, we find a self-consistent equation for $E_P$, the energy of the polaron state
\begin{widetext}
\begin{equation}
E_P-h-\pi\frac{\alpha^2k_F^2}{E_P-2h}=\frac{1}{\cal S}\sum_{q<k_F}\left[\frac{1}{U}-\frac{1}{\cal S}\sum_{k>k_F}\frac{1}{E_P-h-E_{kq}-\frac{2\alpha^2k^2}{E-2h-E_{kq}} -\theta(|\mathbf{q}-\mathbf{k}|-k_F)\frac{2\alpha^2|\mathbf{q}-\mathbf{k}|^2}{E-E_{kq}}}\right]^{-1},\label{eqnpolaronenergy}
\end{equation}
\end{widetext}
where $E_{kq}=\epsilon_{\mathbf{q}-\mathbf{k}}+\epsilon_{\mathbf{k}}-\epsilon_{\mathbf{q}}$, and the third term on the left-hand side corresponds to an energy shift due to the SOC-induced particle-hole fluctuations in the Fermi sea of spin-up atoms. In the weak-coupling limit, this should be the dominant contribution as the summation on the right-hand side of Eq. (\ref{eqnpolaronenergy}) vanishes at $E_b=0$. Hence the polaron energy in the weak-coupling limit is $E_P^f=(3h-\sqrt{h^2+4\pi\alpha^2k_F^2})/2$. Comparing this energy with the molecular energy in the same limit, we can solve for the critical SOC strength of the polaron-molecule transition.

We numerically evaluate the energy of the polaron state, and calculate the critical SOC for the polaron-molecule transition for various $h$ as the interaction is tuned. As is clear from Fig. \ref{figpolaron}, for sufficiently large SOC strength, molecular states are always favored, despite the shift in polaron energy by SOC. In addition, we find that the critical SOC for the polaron-molecule transition is always greater than the critical SOC for the FFLO-like pairing states under the same effective Zeeman field. Therefore at the polaron-molecule transition, the center-of-mass momentum of the molecular state is always $Q=0$. As the molecular state with zero center-of-mass momentum corresponds to the TSF state in the mean field diagram, this polaron-molecule transition implies a phase transition between the TSF state and the normal state for a highly polarized 2D Fermi gas in the thermodynamic limit.

For a 2D Fermi gas, fluctuations are important, and a `bare' molecular state in Eq. (\ref{molstate}) may not be accurate. It has been pointed out in Ref. \cite{parish} that if particle-hole fluctuations are included in the molecular states as has been done for the polaron state, the energy of the molecular state will be lowered. In the presence of SOC, we expect a similar scenario will shift the critical point of polaron-molecule transition to smaller SOC strengths.

\emph{Conclusion}.--
We have studied the novel physics induced by SOC in a 2D Fermi gas in the large polarization limit generated by an effective Zeeman field. With the interplay between SOC, pairing, and particle-hole fluctuations, the system exhibits many interesting properties, e.g. SOC-induced pairing instability, FFLO pairing, polaron-molecule transition, etc. In particular, our results suggest that the particle-hole fluctuations have considerable impact on the stability of the TSF state and can modify the mean-field phase diagram in the large polarization limit. As polaron-molecule transitions in a highly polarized Fermi gas have been experimentally probed recently both in three dimensions \cite{mzpolraon} and in two dimensions \cite{koehlexp}, our results may be tested directly in future experiments.

\emph{Acknowledgements}--
We thank Xiaoling Cui, Ying Ran, Xiaosen Yang , Zengqiang Yu, and Hui Zhai for helpful discussions. This work is supported by NFRP (2011CB921200, 2011CBA00200), NNSF (60921091), NSFC (10904172, 11105134), the Fundamental Research Funds for the Central Universities (WK2470000001, WK2470000006), and the Research Funds of Renmin University of China (10XNL016). W.Z. would also like to thank the NCET program for support.

\newpage

\begin{widetext}
\begin{appendix}
\section{Supplementary material}

In this supplementary material, we present in detail the variational formalism adopted in the main text. We wish to study
the system in the large polarization limit, where a pairing instability persists in the many-body calculations \cite{soc1,wy2d,duan}. In the presence of SOC, the spins are mixed, and the large polarization limit can be achieved by imposing an effective Zeeman field which prevents spin-flipping. The implementation of this effective Zeeman field depends on the specific scheme in realizing the synthetic SOC. For the scheme in Ref. \cite{gauge2exp}  for example, the effective Zeeman field is tunable by adjusting the Rabi-frequency of the Raman lasers. In the canonical ensemble with a total of $N$ particles, the large polarization limit corresponds to the subspace with few spin-down atoms in the presence of a fully-polarized Fermi sea of spin-up atoms. As systems with different population of spin-down atoms can be connected to one another by a simple scaling of the parameters, we further specify the number constraint in the large polarization limit so that there is one spin-down atom on average. To model the ground state in this large polarization limit, we adopt variational ansatz states which effectively project the wave functions into this subspace. As we will see in the following, terms with more than two spin-down atoms are left out of the ansatz, as their contributions are typically small due to the presence of the effective Zeeman field. In practice, the Zeeman field $h$ can be self-consistently determined by imposing the number constraint. Here, instead of fixing the average number of spin-down atoms, we fix the effective Zeeman field $h$, which is of more experimental relevance. The average number of spin-down atoms in our calculation therefore fluctuates around unity.

\subsection{Molecular state}
In this section, we present the formalism for the molecular state. Consider an interacting Fermi gas with SOC and an effective Zeeman field $h$. The Hamiltonian is
\begin{equation}
H=\sum_{\mathbf{k},\sigma}\epsilon_{\mathbf{k}}c^{\dag}_{\mathbf{k},\sigma}c_{\mathbf{k},\sigma}+U\sum_{\mathbf{k},\mathbf{k}',\mathbf{q}} c^{\dag}_{\mathbf{k},\uparrow}c^{\dag}_{\mathbf{k}',\downarrow}c_{\mathbf{k}'+\mathbf{q},\downarrow}c_{\mathbf{k}-\mathbf{q},\uparrow} +\sum_{\mathbf{k}}\left(\alpha k c^{\dag}_{\mathbf{k},\uparrow}c_{\mathbf{k},\downarrow}+\alpha^{\ast}kc^{\dag}_{\mathbf{k},\downarrow}c_{\mathbf{k},\uparrow}\right) +h\sum_{\mathbf{k}}a_{\mathbf{k},\downarrow}^{\dag}a_{\mathbf{k},\downarrow},
\end{equation}
with $\epsilon_{\mathbf{k}}=\hbar^2k^2/2m$. The bare interaction rate $U$ should be renormalized following the standard recipe in 2D
\begin{equation}
\frac{1}{U}=-\frac{1}{\cal S}\sum_{\mathbf{k}}\frac{1}{E_b+2\epsilon_{\mathbf{k}}},
\end{equation}
where $E_b$ is the two-body bound state energy in 2D in the absence of SOC.

For the molecular state, the variational wave function can be written as
\begin{equation}
|\Psi_M\rangle_{\mathbf{Q}}=\sum_{\mathbf{k}}\left(\phi^{\uparrow\downarrow}_{\mathbf{k}}c^{\dag}_{\mathbf{Q}-\mathbf{k},\downarrow}c^{\dag}_{\mathbf{k},\uparrow} +\phi^{\uparrow\uparrow}_{\mathbf{k}}c^{\dag}_{\mathbf{Q}-\mathbf{k},\uparrow}c^{\dag}_{\mathbf{k},\uparrow}+\phi^{\downarrow\downarrow}_{\mathbf{k}} c^{\dag}_{\mathbf{Q}-\mathbf{k},\downarrow}c^{\dag}_{\mathbf{k},\downarrow}\right)|N-1\rangle,
\end{equation}
where $|N-1\rangle$ represents the Fermi sea with $N-1$ spin-up fermions. Importantly, due to the presence of the Fermi sea, the coefficients are constrained such that $k>k_F$ for $\phi^{\uparrow\downarrow}_{\mathbf{k}}$ and $\phi^{\uparrow\uparrow}_{\mathbf{k}}$; and $|\mathbf{Q}-\mathbf{k}|>k_F$ for $\phi^{\uparrow\uparrow}_{\mathbf{k}}$. The normalization condition for the total wave function is
\begin{align}
\langle \Psi_M|\Psi_M\rangle&=\sum_{k>k_F}|\phi^{\uparrow\downarrow}_{\mathbf{k}}|^2+\sum_{k>k_F}\theta(|\mathbf{Q}-\mathbf{k}|-k_F) \left(|\phi^{\uparrow\uparrow}_{\mathbf{k}}|^2-\phi^{\ast\uparrow\uparrow}_{\mathbf{k}}\phi^{\uparrow\uparrow}_{\mathbf{Q}-\mathbf{k}}\right) +\sum_{\mathbf{k}} \left(|\phi^{\downarrow\downarrow}_{\mathbf{k}}|^2-\phi^{\ast\downarrow\downarrow}_{\mathbf{k}}\phi^{\downarrow\downarrow}_{\mathbf{Q}-\mathbf{k}}\right)\nonumber\\
&=1,
\end{align}
where $\theta(x)$ is the Heaviside step function.  Note that the last two terms can be recast into a more symmetric form
\begin{equation}
\sum_{\mathbf{k}}\left(|\phi_{\mathbf{k}}|^2-\phi^{\ast}_{\mathbf{k}}\phi_{\mathbf{Q}-\mathbf{k}}\right)=\sum_{\mathbf{k}} \frac{1}{2}\left(\phi^{\ast}_{\mathbf{k}}-\phi^{\ast}_{\mathbf{Q}-\mathbf{k}}\right)\left(\phi_{\mathbf{k}}-\phi_{\mathbf{Q}-\mathbf{k}}\right).
\end{equation}
We may then evaluate the quantity $\langle \Psi_M|H|\Psi_M\rangle$ and minimize it to derive the equations for the coefficients $\phi_{\mathbf{k}}$ in the wave function.

The expectation value of the energy becomes
\begin{align}
\langle \Psi_M|H|\Psi_M\rangle&=\sum_{k>k_F}|\phi^{\uparrow\downarrow}_{\mathbf{k}}|^2\left(\epsilon_{\mathbf{k}}+\epsilon_{\mathbf{Q}-\mathbf{k}}+ h\right) +\sum_{k>k_F}\frac{1}{2}\left(\phi^{\ast\uparrow\uparrow}_{\mathbf{k}}-\phi^{\ast\uparrow\uparrow}_{\mathbf{Q}-\mathbf{k}}\right) \left(\phi^{\uparrow\uparrow}_{\mathbf{k}}-\phi^{\uparrow\uparrow}_{\mathbf{Q}-\mathbf{k}}\right)\left(\epsilon_{\mathbf{Q}-\mathbf{k}}+\epsilon_{\mathbf{k}}+2h\right) \theta(|\mathbf{Q}-\mathbf{k}|-k_F)\nonumber\\
&+\sum_{\mathbf{k}}\frac{1}{2}\left(\phi^{\ast\downarrow\downarrow}_{\mathbf{k}}-\phi^{\ast\downarrow\downarrow}_{\mathbf{Q}-\mathbf{k}}\right) \left(\phi^{\downarrow\downarrow}_{\mathbf{k}}-\phi^{\downarrow\downarrow}_{\mathbf{Q}-\mathbf{k}}\right)\left(\epsilon_{\mathbf{Q}-\mathbf{k}}+\epsilon_{\mathbf{k}}\right)
+\sum_{k_1>k_F,k_2>k_F}\phi^{\ast\uparrow\downarrow}_{\mathbf{k}_1}\phi^{\uparrow\downarrow}_{\mathbf{k}}U\nonumber\\
&+\left[\sum_{k_1>k_F} \theta(|\mathbf{Q}-\mathbf{k}_1|-k_F)\left(\phi_{\mathbf{k}_1}^{\ast\uparrow\uparrow}-\phi^{\ast\uparrow\uparrow}_{\mathbf{Q}-\mathbf{k}_1}\right)\phi^{\uparrow\downarrow}_{\mathbf{k}_1} \alpha\left|\mathbf{Q}-\mathbf{k}_1\right|+h.c.\right]\nonumber\\
&+\left[\sum_{k_1>k_F}\left(\phi_{\mathbf{k}_1}^{\ast\downarrow\downarrow}-\phi^{\ast\downarrow\downarrow}_{\mathbf{Q}-\mathbf{k}_1}\right)\phi^{\uparrow\downarrow}_{\mathbf{k}_1} \alpha^{\ast}k_1+h.c.\right].
\end{align}

The extremal conditions of the energy functional above then yield the following set of equations for the variational coefficients
\begin{align}
\left(\epsilon_{\mathbf{k}}+\epsilon_{\mathbf{Q}-\mathbf{k}}+h\right)\phi^{\uparrow\downarrow}_{\mathbf{k}}+U\sum_{k'>k_F}\phi^{\uparrow\downarrow}_{\mathbf{k}'} +\alpha^{\ast}\left|\mathbf{Q}-\mathbf{k}\right|\tilde{\phi}^{\uparrow\uparrow}_{\mathbf{k}}+\alpha k\tilde{\phi}^{\downarrow\downarrow}_{\mathbf{k}}&=E\phi^{\uparrow\downarrow}_{\mathbf{k}}\\
\frac{1}{2}\left(\epsilon_{\mathbf{k}}+\epsilon_{\mathbf{Q}-\mathbf{k}}\right)\tilde{\phi}^{\uparrow\uparrow}_{\mathbf{k}}+\alpha\left|\mathbf{Q}-\mathbf{k}\right|\phi^{\uparrow\downarrow}_{\mathbf{k}} &=\frac{1}{2}E\tilde{\phi}^{\uparrow\uparrow}_{\mathbf{k}}\\
\frac{1}{2}\left(\epsilon_{\mathbf{k}}+\epsilon_{\mathbf{Q}-\mathbf{k}}+2h\right)\tilde{\phi}^{\downarrow\downarrow}_{\mathbf{k}}+\alpha^{\ast}k\phi^{\uparrow\downarrow}_{\mathbf{k}} &=\frac{1}{2}E\tilde{\phi}^{\downarrow\downarrow}_{\mathbf{k}},
\end{align}
where we have defined the properly symmetrized coefficients
\begin{align}
\tilde{\phi}_{\mathbf{k}}=\phi_{\mathbf{k}}-\phi_{\mathbf{Q}-\mathbf{k}}.
\end{align}
It is important to keep in mind that in the equations above, we have the constraints $k>k_F$ and $\left|\mathbf{Q}-\mathbf{k}\right|>k_F$. Following the standard procedure, we derive the equations for the ground state energy
\begin{equation}
\frac{1}{U}=\sum_{k>k_F}\left[\frac{1}{E-\left(\epsilon_{\mathbf{k}}+\epsilon_{\mathbf{Q}-\mathbf{k}}+h\right) -\frac{2|\alpha|^2k^2}{E-\left(\epsilon_{\mathbf{k}}+\epsilon_{\mathbf{Q}-\mathbf{k}}+2h\right)} -\theta(|\mathbf{Q}-\mathbf{k}|-k_F)\frac{2|\alpha|^2\left|\mathbf{Q}-\mathbf{k}\right|^2} {E-\left(\epsilon_{\mathbf{k}}+\epsilon_{\mathbf{Q}-\mathbf{k}}\right)}}\right].\label{molenergyeqn}
\end{equation}

Finally, the coefficients can be solved once the ground state energy is fixed
\begin{equation}
\phi^{\uparrow\downarrow}_{\mathbf{k}}=\frac{1}{E-B}\left[\frac{1}{\sum_{k>k_F}\frac{C}{(E-B)^2}}\right]^{\frac{1}{2}},
\end{equation}
where
\begin{align}
&B=\left(\epsilon_{\mathbf{k}}+\epsilon_{\mathbf{Q}-\mathbf{k}}\right)+h+\frac{2|\alpha|^2k^2}{E-2h-\left(\epsilon_{\mathbf{k}} +\epsilon_{\mathbf{Q}-\mathbf{k}}\right)} +\theta(|\mathbf{Q}-\mathbf{k}|-k_F)\frac{2|\alpha|^2\left|\mathbf{Q}-\mathbf{k}\right|^2} {E-\left(\epsilon_{\mathbf{k}}+\epsilon_{\mathbf{Q}-\mathbf{k}}\right)}\\
&C=1+\theta(|\mathbf{Q}-\mathbf{k}|-k_F)\frac{2|\alpha|^2|\mathbf{q}-\mathbf{k}|^2}{(E-\epsilon_{\mathbf{Q}-\mathbf{k}}-\epsilon_{\mathbf{k}})^2} +\frac{2|\alpha|^2k^2}{(E-\epsilon_{\mathbf{Q}-\mathbf{k}}-\epsilon_{\mathbf{k}}-2h)^2}
\end{align}

The average number of spin-down atoms can be calculated from $N_{\downarrow}=\sum_{k>k_F}|\phi^{\uparrow\downarrow}_{\mathbf{k}}|^2+\sum_{\mathbf{k}}|\tilde{\phi}^{\downarrow\downarrow}_{\mathbf{k}}|^2$.
In Fig. \ref{figsupp1}, we illustrate $N_{\downarrow}$ as a function of $E_b$ for fixed SOC strength $\alpha$ and various Zeeman fields. For a fixed Zeeman field with appropriate magnitude, the population of the spin-down atom fluctuates around one, as expected.

\begin{figure}
\includegraphics[width=8cm]{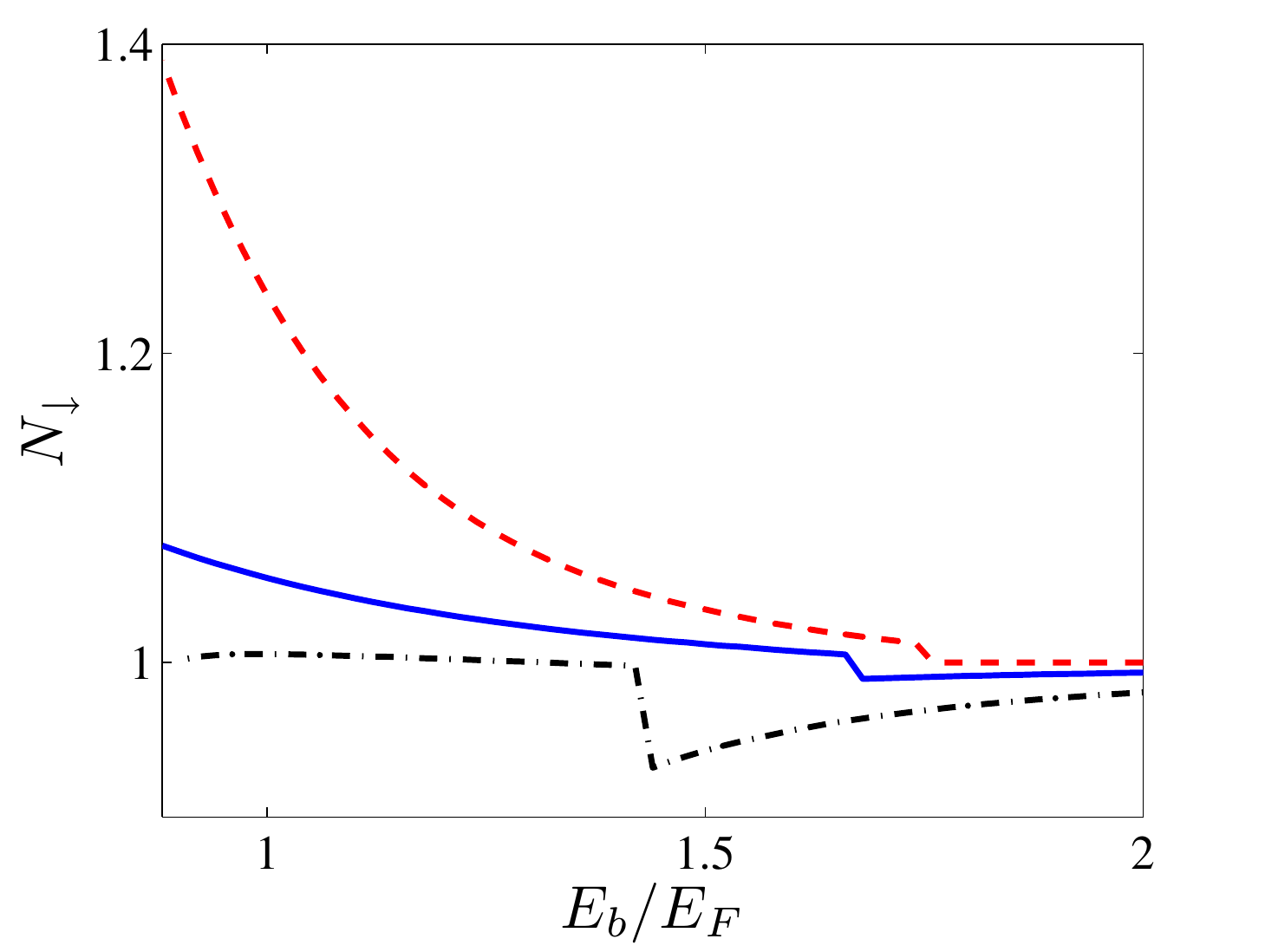}
\caption{(Color online) Average number of spin-down atoms as a function of $E_b$ with fixed SOC strength $\alpha k_F/E_F=0.2$ for various effective Zeeman fields $h$: $h/E_F=0$ (dashed), $h/E_F=0.5$ (solid), $h/E_F=1.1$ (dash-dotted). The abrupt changes indicate first-order transition between the FFLO-like pairing states and pairing states with zero center-of-mass momentum.}
\label{figsupp1}
\end{figure}

\subsection{Polaron state}

Similar to that of the molecular state, the variational wave function for the polaron state can be written as
\begin{align}
|P\rangle&=\psi_0c^{\dag}_{0\downarrow}|N\rangle+\sum_{k>k_F,q<k_F}\psi^{\uparrow\downarrow}_{\mathbf{k},\mathbf{q}}c^{\dag}_{\mathbf{Q}-\mathbf{k},\downarrow}c^{\dag}_{\mathbf{k},\uparrow} c_{\mathbf{q},\uparrow}|N\rangle+\sum_{\mathbf{k},q<k_F}\psi^{\downarrow\downarrow}_{\mathbf{k},\mathbf{q}}c^{\dag}_{\mathbf{q}-\mathbf{k},\downarrow}c^{\dag}_{\mathbf{k},\downarrow} c_{\mathbf{q},\uparrow}|N\rangle\nonumber\\
&+\sum_{q<k_F,k>k_F}\theta(|\mathbf{q}-\mathbf{k}|-k_F)\psi^{\uparrow\uparrow}_{\mathbf{k},\mathbf{q}}c^{\dag}_{\mathbf{q}-\mathbf{k},\uparrow}c^{\dag}_{\mathbf{k},\uparrow}c_{\mathbf{q},\uparrow}|N\rangle.
\end{align}
The normalization condition for the coefficients are
\begin{align}
\langle P|P\rangle
&=|\psi_0|^2+\sum_{k>k_F,q<k_F}\left|\psi^{\uparrow\downarrow}_{\mathbf{k},\mathbf{q}}\right|^2+\sum_{q<k_F,k}\left(\left|\psi^{\downarrow\downarrow}_{\mathbf{k},\mathbf{q}}\right|^2- \psi^{\ast\downarrow\downarrow}_{\mathbf{k},\mathbf{q}}\psi^{\downarrow\downarrow}_{\mathbf{q}-\mathbf{k},\mathbf{q}}\right)\nonumber\\
&+\sum_{q<k_F,k>k_F}\theta(|\mathbf{q}-\mathbf{k}|-k_F)\left(\left|\psi^{\uparrow\uparrow}_{\mathbf{k},\mathbf{q}}\right|^2- \psi^{\ast\uparrow\uparrow}_{\mathbf{k},\mathbf{q}}\psi^{\uparrow\uparrow}_{\mathbf{q}-\mathbf{k},\mathbf{q}}\right)\nonumber\\
&=1.
\end{align}

Minimizing the energy functional in the presence of the Zeeman field, we have
\begin{align}
U\sum_{q<k_F}\psi_0+U\sum_{k>k_F,q<k_F}\psi^{\uparrow\downarrow}_{\mathbf{k},\mathbf{q}}+\sum_{q<k_F}\alpha q\tilde{\psi}^{\downarrow\downarrow}_{\mathbf{q},\mathbf{q}}&=E\psi_0-h\psi_0\\
E_{kq}\psi^{\uparrow\downarrow}_{\mathbf{k},\mathbf{q}}+U\psi_0+U\sum_{k'>k_F}\psi^{\uparrow\downarrow}_{\mathbf{k}',\mathbf{q}}+\alpha k\tilde{\psi}^{\downarrow\downarrow}_{\mathbf{k},\mathbf{q}}+\alpha^{\ast} \left|\mathbf{q}-\mathbf{k}\right|\tilde{\psi}^{\uparrow\uparrow}_{\mathbf{k},\mathbf{q}}&=E\psi^{\uparrow\downarrow}_{\mathbf{k},\mathbf{q}}-h\psi_{\mathbf{k},\mathbf{q}}^{\uparrow\downarrow}\\
\frac{1}{2}E_{kq}\tilde{\psi}^{\downarrow\downarrow}_{\mathbf{k},\mathbf{q}}+\delta_{\mathbf{k},\mathbf{q}}\psi_0\alpha^{\ast}q+\theta(k-k_F)\alpha^{\ast}k\psi^{\uparrow\downarrow}_{\mathbf{k},\mathbf{q}} &=\frac{1}{2}E\tilde{\psi}^{\downarrow\downarrow}_{\mathbf{k},\mathbf{q}}-h\tilde{\psi}^{\downarrow\downarrow}_{\mathbf{k},\mathbf{q}}\\
\frac{1}{2}E_{kq}\tilde{\psi}^{\uparrow\uparrow}_{\mathbf{k},\mathbf{q}}-\alpha k\psi^{\uparrow\downarrow}_{\mathbf{q}-\mathbf{k},\mathbf{q}}&=\frac{1}{2}E\tilde{\psi}^{\uparrow\uparrow}_{\mathbf{k},\mathbf{q}},
\end{align}
where $E_{kq}=\epsilon_{\mathbf{k}}+\epsilon_{\mathbf{q}-\mathbf{k}}-\epsilon_{\mathbf{q}}$, and
\begin{equation}
\tilde{\phi}_{\mathbf{k},\mathbf{q}}=\phi_{\mathbf{k},\mathbf{q}}-\phi_{\mathbf{q}-\mathbf{k},\mathbf{q}}.
\end{equation}

The equation for the energy is
\begin{equation}
E-h-\sum_{q<k_F}\frac{2|\alpha|^2q^2}{E-2h}=\sum_{q<k_F}\left[\frac{1}{U}-\sum_{k>k_F}\frac{1}{E-E_{kq}-h-\frac{|\alpha|^2k^2}{\frac{1}{2}(E-E_{kq}-2h)} -\theta(|\mathbf{q}-\mathbf{k}|-k_F)\frac{|\alpha|^2|\mathbf{q}-\mathbf{k}|^2}{\frac{1}{2}(E-E_{kq})}}\right]^{-1}.
\end{equation}
In the absence of $\alpha$ and $h$ this equation reduces to the corresponding equation for the polaron energy in Ref. \cite{parish}.

From the renormalization condition and the eigen energy equations, we have
\begin{align}
\psi_0&=\left[1+\sum_{k>k_F,q<k_F}C_{kq}\left|\frac{B_{q}}{E-A_{kq}}\right|^2\right]^{-\frac{1}{2}}\nonumber\\
\psi_{\mathbf{k},\mathbf{q}}^{\uparrow\downarrow}&=\frac{B_q}{E-A_{kq}}\psi_0\nonumber\\
\tilde{\psi}_{\mathbf{k},\mathbf{q}}^{\downarrow\downarrow}&=\frac{2\alpha k}{E-E_{kq}-2h}\psi_{\mathbf{k},\mathbf{q}}^{\uparrow\downarrow}\nonumber\\
\tilde{\psi}_{\mathbf{k},\mathbf{q}}^{\uparrow\uparrow}&=\frac{2\alpha |\mathbf{q}-\mathbf{k}|}{E-E_{kq}}\psi_{\mathbf{k},\mathbf{q}}^{\uparrow\downarrow},
\end{align}
where we have defined
\begin{align}
A_{kq}&=E_{kq}+h+\frac{2\alpha^2k^2}{E-E_{kq}-2h}+\theta(|\mathbf{q}-\mathbf{k}|-k_F)\frac{2\alpha^2|\mathbf{q}-\mathbf{k}|^2}{E-E_{kq}}\nonumber\\
B_{q}&=\frac{1}{\frac{1}{U}-\sum_{k'>k_F}\frac{1}{E-A_{k'q}}}\nonumber\\
C_{kq}&=1+\frac{2\alpha^2k^2}{(E-E_{kq}-2h)^2}+\theta(|\mathbf{q}-\mathbf{k}|-k_F)\frac{2\alpha^2|\mathbf{q}-\mathbf{k}|^2}{(E-E_{kq})^2}\nonumber
\end{align}
We have checked numerically that for effective Zeeman fields on the order of $E_F$, the population of the spin-down atoms remains close to unity.

\noindent

\end{appendix}
\end{widetext}

\end{document}